\documentclass[conference]{IEEEtran}
\usepackage{balance}
\usepackage{float}
\usepackage{standard}
\usepackage{etex}

\usepackage{rotate}
\usepackage{algorithmic}
\usepackage{algorithm}
\usepackage{float}
\usepackage{tikz}
\usepackage{refcount}

\usepackage{pgf,tikz}
\usepackage{pgfplots}
\usepackage{pgfplotstable}
\usepackage{bm}
\usepackage{adjustbox}
\usetikzlibrary{calc}
\usepackage{relsize}
\usepackage{ifthen}

\usetikzlibrary{calc,fit,arrows,plotmarks,intersections,patterns,shapes,decorations,positioning, backgrounds, matrix}

\usepackage[numbers,sort&compress]{natbib}
\usepackage[ngerman, english]{babel}

\usetikzlibrary{arrows.meta}
\edef\wdArrowLength{2}
\tikzset{>={Latex[width=1.5mm,length=\wdArrowLength mm]}}

\usepackage{cancel}
\usepackage{mathtools}
\usepackage{shadethm}
\usepackage{empheq}
\usepackage{skull}
\usepgfplotslibrary{units}
\usepackage{calc}
\usepackage{nicefrac}
\usepackage{fancybox}
\usepackage{psfrag}
\usepackage{dsfont}

\title{Dynamic Rate Splitting Grouping for Antifragile Responses to Wireless Network Disruptions
} 
\author{\IEEEauthorblockN{Kevin Weinberger, Aydin Sezgin}
\IEEEauthorblockA{Ruhr-University Bochum, Germany\\
Email: \{kevin.weinberger,aydin.sezgin\}@rub.de}
\thanks{This work was supported in part by the German Federal Ministry of Education and Research (BMBF) in the course of the 6GEM Research Hub under grant 16KISK03.}
}
\date{\today}
\usepackage{graphicx}
\usepackage{placeins}

\usepackage{booktabs}
\usepackage{marvosym}
\usepackage{multirow}

\usepackage{latexsym, amsmath, amssymb, amsfonts, upgreek}

\usepackage[nolist]{acronym}

\usepackage{tikz}
\usetikzlibrary{calc,arrows,positioning,decorations,shadows,shapes,fadings,matrix}
\tikzset{>=latex'}
\tikzset{semithick}

\makeatletter
\providecommand{\IfElsePackageLoaded}[3]{\@ifpackageloaded{#1}{#2}{#3}}
\makeatother

\usepackage{subfigure}
\IfElsePackageLoaded{subfig}
	{\usepackage[subfigure]{tocloft}}{	
	\IfElsePackageLoaded{subfigure}
		{\usepackage[subfigure]{tocloft}}
		{\usepackage{tocloft}}
	}

\makeatletter

\def\tikz@delimiter#1#2#3#4#5#6#7#8{%
	\bgroup
		\pgfextra{\let\tikz@save@last@fig@name=\tikz@last@fig@name}%
		node[outer sep=0pt,inner sep=0pt,draw=none,fill=none,anchor=#1,at=(\tikz@last@fig@name.#2),#3]
		{%
			{\nullfont\pgf@process{\pgfpointdiff{\pgfpointanchor{\tikz@last@fig@name}{#4}}{\pgfpointanchor{\tikz@last@fig@name}{#5}}}}%
			\delimitershortfall\z@
			\resizebox*{!}{#8}{$\left#6\vcenter{\hrule height .5#8 depth .5#8 width0pt}\right#7$}%
		}
		\pgfextra{\global\let\tikz@last@fig@name=\tikz@save@last@fig@name}%
	\egroup%
}

\newcommand{\skewedIRS}{
	\begin{tikzpicture}
		\node[rectangle,minimum width=1.5cm,minimum height=1.5cm,draw,yslant=-0.6,scale=0.5,fill=blue!20!green] at (0,0){};

		\node[rectangle,minimum width=0.05cm,minimum height=0.05cm,draw,yslant=-0.6,scale=0.25,fill=yellow!70!green] (base) at (-0.3,-0.12){};

         \foreach \y in {1,...,5}
         {
            \node[rectangle,minimum width=0.05cm,minimum height=0.05cm,draw,yslant=-0.6,scale=0.25,fill=yellow!70!green] at
            ($($(base)+(-31:0.14*0)$)+(90:0.12*\y)$){};
         }

      \foreach \x in {1,...,5}
      {
         \node[rectangle,minimum width=0.05cm,minimum height=0.05cm,draw,yslant=-0.6,scale=0.25,fill=yellow!70!green] at
         ($(base)+(-31:0.14*\x)$){};

         \foreach \y in {1,...,5}
         {
            \node[rectangle,minimum width=0.05cm,minimum height=0.05cm,draw,yslant=-0.6,scale=0.25,fill=yellow!70!green] at
            ($($(base)+(-31:0.14*\x)$)+(90:0.12*\y)$){};
         }
      }
	\end{tikzpicture}	
}
\tikzset{hexagon/.code={
	\draw (0,2) -- (-4,0) -- (0,-2) -- (4,0) -- (0,2);
}}

\tikzset{phone/.code={
   \node [rectangle,rounded corners=1.5pt,draw,minimum height=0.6cm, minimum width=0.35cm] at (0,0){};
   \node [rectangle,rounded corners=1.5pt,draw,minimum height=0.5cm, minimum width=0.3cm] at (0,0){};
}}

\makeatletter
\def\cantox@vector#1#2#3#4#5#6#7#8{%
  \dimen@.5\p@
  \setbox\z@\vbox{\boxmaxdepth.5\p@
   \hbox{\kern-1.2\p@\kern#1\dimen@$#7{#8}\m@th$}}%
  \ifx\canto@fil\hidewidth  \wd\z@\z@ \else \kern-#6\unitlength \fi
  \ooalign{%
    \canto@fil$\m@th \CancelColor
    \vcenter{\hbox{\dimen@#6\unitlength \kern\dimen@
      \multiply\dimen@#4\divide\dimen@#3 \vrule\@depth\dimen@\@width\z@
      \vector(#3,-#4){#5}%
    }}_{\raise-#2\dimen@\copy\z@\kern-\scriptspace}$%
    \canto@fil \cr
    \hfil \box\@tempboxa \kern\wd\z@ \hfil \cr}}
\def\bcancelto#1#2{\let\canto@vector\cantox@vector\cancelto{#1}{#2}}
\makeatother

\newcommand{\ifthen}[2]{\ifthenelse{#1}{#2}{}}

\newcommand{\sbrackets}[1]{\left(#1\right)}
\newcommand{\hbrackets}[1]{\left[#1\right]}
\newcommand{\vectw}{\vect{\omega}}

\newcommand{\vecth}{\vect{h}}

\newcommand{\cbrackets}[1]{\left\{#1\right\}}

\newcommand{\IRS}{
	\begin{tikzpicture}
		\node[rectangle,minimum width=1.5cm,minimum height=1.5cm,draw,scale=0.5,fill=blue!20!green] at (0,0){};

		\node[rectangle,minimum width=0.05cm,minimum height=0.05cm,draw,scale=0.25,fill=yellow!70!green] (base) at (-0.3,-0.3){};

         \foreach \y in {1,...,5}
         {
            \node[rectangle,minimum width=0.05cm,minimum height=0.05cm,draw,scale=0.25,fill=yellow!70!green] at
            ($($(base)+(-31:0.14*0)$)+(90:0.12*\y)$){};
         }

      \foreach \x in {1,...,5}
      {
         \node[rectangle,minimum width=0.05cm,minimum height=0.05cm,draw,scale=0.25,fill=yellow!70!green] at
         ($(base)+(0:0.12*\x)$){};

         \foreach \y in {1,...,5}
         {
            \node[rectangle,minimum width=0.05cm,minimum height=0.05cm,draw,scale=0.25,fill=yellow!70!green] at
            ($($(base)+(0:0.12*\x)$)+(90:0.12*\y)$){};
         }
      }
	\end{tikzpicture}	
}

\newcommand{\listequationsname}{List of Formulas}
\newlistof{myequations}{equ}{\listequationsname}

\definecolor{myblue1}{rgb}{0,0,255}
\definecolor{myblue2}{rgb}{65,105,225}
\definecolor{myblue3}{rgb}{70,130,180}
\definecolor{myblue4}{rgb}{176,196,222}

\newcommand{\mytilde}{{\raise.17ex\hbox{$\scriptstyle\mathtt{\sim}$}}}
\newcommand{\naive}{}
\def\naive/{na\"{\i}ve}

\newcommand{\executeiffilenewer}[3]{%
\ifnum\pdfstrcmp{\pdffilemoddate{#1}}%
{\pdffilemoddate{#2}}>0%
{\immediate\write18{#3}}\fi%
}

\usepackage{tikz}
\usetikzlibrary{arrows}
\usetikzlibrary{patterns}
\tikzset{>=latex'}
\usetikzlibrary{shapes.geometric}
\usetikzlibrary{calc}
\usetikzlibrary{positioning, shapes,fadings,through}

\pgfkeys{/pgf/fpu}
\pgfmathparse{16383+1}
\edef\tmp{\pgfmathresult}
\pgfkeys{/pgf/fpu=false}
\newcommand{\iPhone}[3]{
	\coordinate (a) at (#1,#2);
	\draw [line width=0.25pt,rounded corners=(#3)*1mm,fill=white,scale=(#3)] (a)--($(a)+(0.67,0)$)--($(a)+(0.67,1.381)$)--($(a)+(0,1.381)$)--cycle;
	\draw [color=gray,line width=0.25pt,rounded corners=(#3)*0.8mm,fill=white,scale=(#3)] ($(a)+(0.015,0.015)$)--($(a)+(0.655,0.015)$)--($(a)+(0.655,1.366)$)--($(a)+(0.015,1.366)$)--cycle;
	\draw [line width=0.25pt,rounded corners=(#3)*0.04mm,scale=(#3)] ($(a)+(0.2875,1.266)$)--($(a)+(0.3825,1.266)$)--($(a)+(0.3825,1.281)$)--($(a)+(0.2875,1.281)$)--cycle;
	\draw[line width=0.25pt,scale=#3] ($(a)+(0.335,0.09)$) circle (0.055cm);
	\draw[line width=0.25pt,scale=#3] ($(a)+(0.335,0.09)$) circle (0.044cm);
	\draw[line width=0.25pt,scale=#3] ($(a)+(0.2275,1.2735)$) circle (0.015cm);
	\draw[line width=0.25pt,scale=#3] ($(a)+(0.335,1.32)$) circle (0.01cm);
	\draw [fill={rgb:black,1;white,4},line width=0.25pt,scale=(#3)] ($(a)+(0.042475,0.170195)$)--($(a)+(0.042475,0.170195)+(0.58505,0.0)$)--($(a)+(0.042475,0.170195)+(0.58505,1.04061)$)--($(a)+(0.042475,0.170195)+(0.0,1.04061)$)--cycle;
}
\newcommand{\basestation}[3]{
	\coordinate (a) at (#1,#2);
	\draw[line width=(#3)*1.5pt,scale=#3] ($(a)+(0, -1.08)$) -- ($(a)+(0, 1)$);
	\draw[line width=(#3)*1.5pt,scale=#3] ($(a)+(0,1)$) .. controls ($(a)+(-0.25,-0.5)$) .. ($(a)+(-0.5,-0.9)$);
	\draw[line width=(#3)*1.5pt,scale=#3] ($(a)+(0,1)$) .. controls ($(a)+(0.25,-0.5)$) .. ($(a)+(0.5,-0.9)$);
	\draw[line width=(#3)*1.5pt,scale=#3] ($(a)+(0, -0.9)$) -- ($(a)+(-0.35, -0.7)$);
	\draw[line width=(#3)*1.5pt,scale=#3] ($(a)+(0, -0.9)$) -- ($(a)+(0.35, -0.7)$);
	\draw[line width=(#3)*0.75pt,scale=#3] ($(a)+(-0.35, -0.65)$) -- ($(a)+(0, -0.5)$);
	\draw[line width=(#3)*0.75pt,scale=#3] ($(a)+(0.35, -0.65)$) -- ($(a)+(0, -0.5)$);
	\draw[line width=(#3)*1pt,scale=#3] ($(a)+(0, -0.6)$) -- ($(a)+(-0.3, -0.45)$);
	\draw[line width=(#3)*1pt,scale=#3] ($(a)+(0, -0.6)$) -- ($(a)+(0.3, -0.45)$);
	\draw[line width=(#3)*0.5pt,scale=#3] ($(a)+(-0.3, -0.45)$) -- ($(a)+(0, -0.32)$);
	\draw[line width=(#3)*0.5pt,scale=#3] ($(a)+(0.3, -0.45)$) -- ($(a)+(0, -0.32)$);
	\draw[line width=(#3)*0.75pt,scale=#3] ($(a)+(0, -0.3)$) -- ($(a)+(-0.22, -0.17)$);
	\draw[line width=(#3)*0.75pt,scale=#3] ($(a)+(0, -0.3)$) -- ($(a)+(0.22, -0.17)$);
	\draw[line width=(#3)*0.5pt,scale=#3] ($(a)+(-0.22, -0.17)$) -- ($(a)+(0, -0.07)$);
	\draw[line width=(#3)*0.5pt,scale=#3] ($(a)+(0.22, -0.17)$) -- ($(a)+(0, -0.07)$);;
	\draw[line width=(#3)*0.75pt,scale=#3] (a) -- ($(a)+(-0.18, 0.11)$);
	\draw[line width=(#3)*0.75pt,scale=#3] (a) -- ($(a)+(0.18, 0.11)$);
	\draw[line width=(#3)*0.5pt,scale=#3] ($(a)+(-0.18, 0.11)$) -- ($(a)+(0,0.2)$);
	\draw[line width=(#3)*0.5pt,scale=#3] ($(a)+(0.18, 0.11)$) -- ($(a)+(0,0.2)$);
	\draw[line width=(#3)*0.5pt,scale=#3] ($(a)+(0, 0.3)$) -- ($(a)+(-0.1, 0.37)$);
	\draw[line width=(#3)*0.5pt,scale=#3] ($(a)+(0, 0.3)$) -- ($(a)+(0.1, 0.37)$);
	\draw[line width=(#3)*0.25pt,scale=#3] ($(a)+(-0.1, 0.37)$) -- ($(a)+(0, 0.43)$);
	\draw[line width=(#3)*0.25pt,scale=#3] ($(a)+(0.1, 0.37)$) -- ($(a)+(0, 0.43)$);
	\draw[line width=(#3)*0.75pt,scale=#3] ($(a)+(0, 1.2)$) -- ($(a)+(0,1)$);
	\draw[fill=white,scale=#3] ($(a)+(0, 1.2)$) circle (0.05cm);
	\draw[line width=(#3)*1pt,decorate,decoration=expanding waves,decoration={segment length=(#3)*10pt},scale=#3] ($(a)+(0, 1.2)$) -- ($(a)+(0, 2.5)$);
}
\newcommand{\basestationtwoant}[3]{
	\coordinate (a) at (#1,#2);
	\draw[line width=(#3)*1.5pt,scale=#3] ($(a)+(0, -1.08)$) -- ($(a)+(0, 1)$);
	\draw[line width=(#3)*1.5pt,scale=#3] ($(a)+(0,1)$) .. controls ($(a)+(-0.25,-0.5)$) .. ($(a)+(-0.5,-0.9)$);
	\draw[line width=(#3)*1.5pt,scale=#3] ($(a)+(0,1)$) .. controls ($(a)+(0.25,-0.5)$) .. ($(a)+(0.5,-0.9)$);
	\draw[line width=(#3)*1.5pt,scale=#3] ($(a)+(0, -0.9)$) -- ($(a)+(-0.35, -0.7)$);
	\draw[line width=(#3)*1.5pt,scale=#3] ($(a)+(0, -0.9)$) -- ($(a)+(0.35, -0.7)$);
	\draw[line width=(#3)*0.75pt,scale=#3] ($(a)+(-0.35, -0.65)$) -- ($(a)+(0, -0.5)$);
	\draw[line width=(#3)*0.75pt,scale=#3] ($(a)+(0.35, -0.65)$) -- ($(a)+(0, -0.5)$);
	\draw[line width=(#3)*1pt,scale=#3] ($(a)+(0, -0.6)$) -- ($(a)+(-0.3, -0.45)$);
	\draw[line width=(#3)*1pt,scale=#3] ($(a)+(0, -0.6)$) -- ($(a)+(0.3, -0.45)$);
	\draw[line width=(#3)*0.5pt,scale=#3] ($(a)+(-0.3, -0.45)$) -- ($(a)+(0, -0.32)$);
	\draw[line width=(#3)*0.5pt,scale=#3] ($(a)+(0.3, -0.45)$) -- ($(a)+(0, -0.32)$);
	\draw[line width=(#3)*0.75pt,scale=#3] ($(a)+(0, -0.3)$) -- ($(a)+(-0.22, -0.17)$);
	\draw[line width=(#3)*0.75pt,scale=#3] ($(a)+(0, -0.3)$) -- ($(a)+(0.22, -0.17)$);
	\draw[line width=(#3)*0.5pt,scale=#3] ($(a)+(-0.22, -0.17)$) -- ($(a)+(0, -0.07)$);
	\draw[line width=(#3)*0.5pt,scale=#3] ($(a)+(0.22, -0.17)$) -- ($(a)+(0, -0.07)$);;
	\draw[line width=(#3)*0.75pt,scale=#3] (a) -- ($(a)+(-0.18, 0.11)$);
	\draw[line width=(#3)*0.75pt,scale=#3] (a) -- ($(a)+(0.18, 0.11)$);
	\draw[line width=(#3)*0.5pt,scale=#3] ($(a)+(-0.18, 0.11)$) -- ($(a)+(0,0.2)$);
	\draw[line width=(#3)*0.5pt,scale=#3] ($(a)+(0.18, 0.11)$) -- ($(a)+(0,0.2)$);
	\draw[line width=(#3)*0.5pt,scale=#3] ($(a)+(0, 0.3)$) -- ($(a)+(-0.1, 0.37)$);
	\draw[line width=(#3)*0.5pt,scale=#3] ($(a)+(0, 0.3)$) -- ($(a)+(0.1, 0.37)$);
	\draw[line width=(#3)*0.25pt,scale=#3] ($(a)+(-0.1, 0.37)$) -- ($(a)+(0, 0.43)$);
	\draw[line width=(#3)*0.25pt,scale=#3] ($(a)+(0.1, 0.37)$) -- ($(a)+(0, 0.43)$);
	\draw[line width=(#3)*1pt,decorate,decoration=expanding waves,decoration={segment length=(#3)*10pt},scale=#3] ($(a)+(0, 1.2)$) -- ($(a)+(0, 2.5)$);
	\draw[line width=(#3)*0.75pt,scale=#3] ($(a)+(0.2, 1)$) -- ($(a)+(0,1)$);
	\draw[line width=(#3)*0.75pt,scale=#3] ($(a)+(0.2, 1.2)$) -- ($(a)+(0.2,1)$);
	\draw[fill=white,scale=#3] ($(a)+(0.2, 1.2)$) circle (0.05cm);
	\draw[line width=(#3)*0.75pt,scale=#3] ($(a)+(-0.2, 1)$) -- ($(a)+(0,1)$);
	\draw[line width=(#3)*0.75pt,scale=#3] ($(a)+(-0.2, 1.2)$) -- ($(a)+(-0.2,1)$);
	\draw[fill=white,scale=#3] ($(a)+(-0.2, 1.2)$) circle (0.05cm);
}
\newcommand{\basestationtwoantno}[3]{
	\coordinate (a) at (#1,#2);
	\draw[line width=(#3)*1.5pt,scale=#3] ($(a)+(0, -1.08)$) -- ($(a)+(0, 1)$);
	\draw[line width=(#3)*1.5pt,scale=#3] ($(a)+(0,1)$) .. controls ($(a)+(-0.25,-0.5)$) .. ($(a)+(-0.5,-0.9)$);
	\draw[line width=(#3)*1.5pt,scale=#3] ($(a)+(0,1)$) .. controls ($(a)+(0.25,-0.5)$) .. ($(a)+(0.5,-0.9)$);
	\draw[line width=(#3)*1.5pt,scale=#3] ($(a)+(0, -0.9)$) -- ($(a)+(-0.35, -0.7)$);
	\draw[line width=(#3)*1.5pt,scale=#3] ($(a)+(0, -0.9)$) -- ($(a)+(0.35, -0.7)$);
	\draw[line width=(#3)*0.75pt,scale=#3] ($(a)+(-0.35, -0.65)$) -- ($(a)+(0, -0.5)$);
	\draw[line width=(#3)*0.75pt,scale=#3] ($(a)+(0.35, -0.65)$) -- ($(a)+(0, -0.5)$);
	\draw[line width=(#3)*1pt,scale=#3] ($(a)+(0, -0.6)$) -- ($(a)+(-0.3, -0.45)$);
	\draw[line width=(#3)*1pt,scale=#3] ($(a)+(0, -0.6)$) -- ($(a)+(0.3, -0.45)$);
	\draw[line width=(#3)*0.5pt,scale=#3] ($(a)+(-0.3, -0.45)$) -- ($(a)+(0, -0.32)$);
	\draw[line width=(#3)*0.5pt,scale=#3] ($(a)+(0.3, -0.45)$) -- ($(a)+(0, -0.32)$);
	\draw[line width=(#3)*0.75pt,scale=#3] ($(a)+(0, -0.3)$) -- ($(a)+(-0.22, -0.17)$);
	\draw[line width=(#3)*0.75pt,scale=#3] ($(a)+(0, -0.3)$) -- ($(a)+(0.22, -0.17)$);
	\draw[line width=(#3)*0.5pt,scale=#3] ($(a)+(-0.22, -0.17)$) -- ($(a)+(0, -0.07)$);
	\draw[line width=(#3)*0.5pt,scale=#3] ($(a)+(0.22, -0.17)$) -- ($(a)+(0, -0.07)$);;
	\draw[line width=(#3)*0.75pt,scale=#3] (a) -- ($(a)+(-0.18, 0.11)$);
	\draw[line width=(#3)*0.75pt,scale=#3] (a) -- ($(a)+(0.18, 0.11)$);
	\draw[line width=(#3)*0.5pt,scale=#3] ($(a)+(-0.18, 0.11)$) -- ($(a)+(0,0.2)$);
	\draw[line width=(#3)*0.5pt,scale=#3] ($(a)+(0.18, 0.11)$) -- ($(a)+(0,0.2)$);
	\draw[line width=(#3)*0.5pt,scale=#3] ($(a)+(0, 0.3)$) -- ($(a)+(-0.1, 0.37)$);
	\draw[line width=(#3)*0.5pt,scale=#3] ($(a)+(0, 0.3)$) -- ($(a)+(0.1, 0.37)$);
	\draw[line width=(#3)*0.25pt,scale=#3] ($(a)+(-0.1, 0.37)$) -- ($(a)+(0, 0.43)$);
	\draw[line width=(#3)*0.25pt,scale=#3] ($(a)+(0.1, 0.37)$) -- ($(a)+(0, 0.43)$);
	\draw[line width=(#3)*0.75pt,scale=#3] ($(a)+(0.2, 1)$) -- ($(a)+(0,1)$);
	\draw[line width=(#3)*0.75pt,scale=#3] ($(a)+(0.2, 1.2)$) -- ($(a)+(0.2,1)$);
	\draw[fill=white,scale=#3] ($(a)+(0.2, 1.2)$) circle (0.05cm);
	\draw[line width=(#3)*0.75pt,scale=#3] ($(a)+(-0.2, 1)$) -- ($(a)+(0,1)$);
	\draw[line width=(#3)*0.75pt,scale=#3] ($(a)+(-0.2, 1.2)$) -- ($(a)+(-0.2,1)$);
	\draw[fill=white,scale=#3] ($(a)+(-0.2, 1.2)$) circle (0.05cm);
}
\renewcommand{\skewedIRS}{
	\begin{tikzpicture}
		\node[rectangle,minimum width=1.5cm,minimum height=1.5cm,draw,yslant=-0.6,scale=0.5,fill=blue!20!green] at (0,0){};
		
		\node[rectangle,minimum width=0.05cm,minimum height=0.05cm,draw,yslant=-0.6,scale=0.25,fill=yellow!70!green] (base) at (-0.3,-0.12){};
		
		\foreach \y in {1,...,5}
		{
			\node[rectangle,minimum width=0.05cm,minimum height=0.05cm,draw,yslant=-0.6,scale=0.25,fill=yellow!70!green] at
			($($(base)+(-31:0.14*0)$)+(90:0.12*\y)$){};
		}
		
		\foreach \x in {1,...,5}
		{
			\node[rectangle,minimum width=0.05cm,minimum height=0.05cm,draw,yslant=-0.6,scale=0.25,fill=yellow!70!green] at
			($(base)+(-31:0.14*\x)$){};
			
			\foreach \y in {1,...,5}
			{
				\node[rectangle,minimum width=0.05cm,minimum height=0.05cm,draw,yslant=-0.6,scale=0.25,fill=yellow!70!green] at
				($($(base)+(-31:0.14*\x)$)+(90:0.12*\y)$){};
			}
		}
	\end{tikzpicture}	
}

\renewcommand{\IRS}{
	\begin{tikzpicture}
		\node[rectangle,minimum width=1.5cm,minimum height=1.5cm,draw,scale=0.5,fill=blue!20!green] at (0,0){};
		
		\node[rectangle,minimum width=0.05cm,minimum height=0.05cm,draw,scale=0.25,fill=yellow!70!green] (base) at (-0.3,-0.3){};
		
		\foreach \y in {1,...,5}
		{
			\node[rectangle,minimum width=0.05cm,minimum height=0.05cm,draw,scale=0.25,fill=yellow!70!green] at
			($($(base)+(-31:0.14*0)$)+(90:0.12*\y)$){};
		}
		
		\foreach \x in {1,...,5}
		{
			\node[rectangle,minimum width=0.05cm,minimum height=0.05cm,draw,scale=0.25,fill=yellow!70!green] at
			($(base)+(0:0.12*\x)$){};
			
			\foreach \y in {1,...,5}
			{
				\node[rectangle,minimum width=0.05cm,minimum height=0.05cm,draw,scale=0.25,fill=yellow!70!green] at
				($($(base)+(0:0.12*\x)$)+(90:0.12*\y)$){};
			}
		}
	\end{tikzpicture}	
}
\newcommand{\tiltedIRS}{
	\begin{tikzpicture}
		\node[rectangle,minimum width=1.5cm,minimum height=1.5cm,draw,scale=0.5,fill=blue!20!green] at (0,0){};
		
		\node[rectangle,minimum width=0.05cm,minimum height=0.05cm,draw,scale=0.25,fill=yellow!70!green] (base) at (-0.3,-0.3){};
		
		\foreach \y in {1,...,5}
		{
			\node[rectangle,minimum width=0.05cm,minimum height=0.05cm,draw,scale=0.25,fill=yellow!70!green] at
			($($(base)+(-31:0.14*0)$)+(90:0.12*\y)$){};
		}
		
		\foreach \x in {1,...,5}
		{
			\node[rectangle,minimum width=0.05cm,minimum height=0.05cm,draw,scale=0.25,fill=yellow!70!green] at
			($(base)+(0:0.12*\x)$){};
			
			\foreach \y in {1,...,5}
			{
				\node[rectangle,minimum width=0.05cm,minimum height=0.05cm,draw,scale=0.25,fill=yellow!70!green] at
				($($(base)+(0:0.12*\x)$)+(90:0.12*\y)$){};
			}
		}
	\end{tikzpicture}	
}

\newcommand{\sensor}[4]{
	\coordinate (a) at (#1,#2);
	\draw[#4,fill=#4,scale=#3,rounded corners=1.5pt] ($(a)+(-.4, -.1)$) rectangle ++(.8,-.3);
	\draw[#4,fill=#4,scale=#3] ($(a)+(-.4, 0)$) rectangle ++(.8,-.18);
	\draw[white,fill=white,scale=#3] ($(a)+(0, 0)$) circle (0.2cm);
	\draw[#4,fill=#4,scale=#3] ($(a)+(0, 0)$) circle (0.15cm);
	\draw[#4,line width=(#3)*1pt,decorate,decoration=expanding waves,decoration={segment length=(#3)*10pt},scale=#3] ($(a)+(0, 0)$) -- ($(a)+(0, 1.3)$);
}
\newcommand{\criticality}[4]{
	\coordinate (a) at (#1,#2);
	\draw[#4,scale=#3,rounded corners=1.5pt] ($(a)+(0,0)$) rectangle ++(.8,-.8);
	\node[#4,scale=#3,anchor=center,align=center] at ($(a)+(.14,-.15)$) {\LARGE\bfseries{!}};
	\draw[#4,scale=#3,rounded corners=1pt] ($(a)+(.05,-.7)$) -- ($(a)+(.4, -.05)$) -- ($(a)+(.75, -.7)$) -- cycle;
}
\newcommand{\actuator}[4]{
	\coordinate (a) at (#1,#2);
	\draw[#4,scale=#3] ($(a)+(-.2, .6)$) rectangle ++(.4,.41);
	\draw[#4,scale=#3] ($(a)+(0,0)$) -- ($(a)+(0, .6)$);
	\node[#4,anchor=center,align=center] at ($(a)+(0,.325)$) {\tiny A};
	\draw[#4,scale=#3] ($(a)+(0,0)$) -- ($(a)+(.9, .6)$) -- ($(a)+(.9, -.6)$) -- cycle;
	\draw[#4,scale=#3] ($(a)+(0,0)$) -- ($(a)+(-.9, .6)$) -- ($(a)+(-.9, -.6)$) -- cycle;
	\draw[#4,fill=#4!35,scale=#3] ($(a)+(0, 0)$) circle (0.3cm);	
}
\newcommand{\ap}[4]{
	\coordinate (a) at (#1,#2);
	\draw[white,fill=#4!40,scale=#3] ($(a)+(0, 0)$) circle (0.25cm);
	\draw[#4,fill=white,scale=#3] ($(a)+(0, 0)$) circle (0.15cm);
	\draw[#4,scale=#3] ($(a)+(0, 0)$) circle (0.4cm);
	\draw[#4,line width=(#3)*0.75pt,decorate,decoration=expanding waves,rotate=0,decoration={segment length=(#3)*5pt},scale=#3,yshift=-10pt] ($(a)+(.4, 0)$) -- ($(a)+(.8, 0)$);
	\draw[#4,line width=(#3)*0.75pt,decorate,decoration=expanding waves,rotate=0,decoration={segment length=(#3)*5pt},scale=#3,yshift=-10pt] ($(a)+(-.4, 0)$) -- ($(a)+(-.8, 0)$);
}

\usetikzlibrary{decorations}
\pgfdeclaredecoration{lightning bolt}{draw}{
	\state{draw}[width=\pgfdecoratedpathlength]{
		\pgfpathmoveto{\pgfpointorigin}%
		\pgfpathlineto{\pgfpoint{\pgfdecoratedpathlength*0.6}%
			{-\pgfdecoratedpathlength*.1}}%
		\pgfpathlineto{\pgfpoint{\pgfdecoratedpathlength*0.55}{0pt}}%
		\pgfpathlineto{\pgfpoint{\pgfdecoratedpathlength}{0pt}}%
		\pgfpathlineto{\pgfpoint{\pgfdecoratedpathlength*0.4}%
			{\pgfdecoratedpathlength*.1}}%
		\pgfpathlineto{\pgfpoint{\pgfdecoratedpathlength*0.45}{0pt}}%
		\pgfpathclose%
	}%
}

\IEEEoverridecommandlockouts
\begin{document}
\maketitle
\begin{abstract}
The reliance on wireless network architectures for applications demanding high reliability and fault tolerance is growing. These architectures heavily depend on wireless channels, making them susceptible to impairments and blockages. Ensuring functionality, particularly for safety-critical applications, demands robust countermeasures at the physical layer. In response, this work proposes the utilization of a dynamic Rate Splitting (RS) grouping approach as a resilience mechanism during blockages. RS effectively manages interference within networks but faces challenges during outages and blockages, where system performance can deteriorate due to the lowest decoding rate dictating the common rate and increased interference from fewer available channel links. As a strategic countermeasure, RS is leveraged to mitigate the impact of blockages, maintaining system efficiency and performance amidst disruptions. In fact, the introduction of new RS groups enables the exploration of novel solutions to the resource allocation problem, potentially outperforming those adopted before the occurrence of a blockage. As it turns out, by employing the dynamic RS grouping, the network exhibits an antifragile recovery response, showcasing the network's ability to not only recover from disruptions but also surpass its initial performance. \vspace{-0.015cm}
\end{abstract} 
\thispagestyle{empty}
\pagestyle{empty}

\section{Introduction}
%
Enterprises increasingly adopt next-gen low-power wide-area access tech for \ac{IoT} networks, offering better energy efficiency, reliability, and higher capacities. This shift is crucial for agile operations in sectors like gigafactories and green steel plants, where cables are insufficient. \cite{ericsson}. Such applications demand \acp{URLLC}, forming the backbone of operations reliant on continuous data streams \cite{crit3}. Consequently, the wireless communication network shoulders an increased responsibility. However, the inherent unpredictability of the wireless channel, coupled with shadowing and blockages, can lead to outages and failed packet deliveries. Depending on the severity, outages can range from minor delays to jeopardizing safety and environmental concerns, necessitating the network's ability to evaluate and respond to such events while upholding service standards \cite{resiliencemetric,RobertRes,RIS_Res}.

Challenges in \ac{URLLC} systems include the inability to use legacy retransmission methods for failed packet deliveries. Instead, continuous network monitoring and swift implementation of resilience mechanisms are necessary to counteract potential outages \cite{resiliencemetric}. These mechanisms involve reallocating network resources as needed \cite{RobertRes,RIS_Res}.

To enhance resilience in URLLC systems, \ac{RS} has emerged as a promising technique \cite{RobertRes}. However, conventional RS primarily focuses on minimizing interference, which can lead to situations where all users decode multiple common messages. Consequently, the resulting resource allocation becomes more rigid, requiring all users within the RS group to decode at the same rate. This rigidity limits the network's ability to adapt to changes, as disruptions to one user's channel may affect decoding for all users within the groups associated with the affected user.
Furthermore, these \ac{RS} groups are established without prior awareness of the network's disruption, potentially rendering them suboptimal when blockages occur. Modifying the interconnected RS groups post-disruption poses the risk of hindering decoding for all users within a group. This can lead to significantly degraded performance and underscores the trade-off between efficiency and flexibility in resilient systems.

In this work, we utilize the fact that the network will reconfigure itself to counteract a loss in performance every time a blockage occurs. Our aim is to harness the concept of antifragility, which is fundamentally different from resilience as it involves actively capitalizing on volatilities, i.e., changes, in the network \cite{AF_wirel,barbell,barbell2,Antifrag_Tip}. Antifragility spans various concepts across different scientific domains (refer to \cite{barbell,barbell2,Antifrag_Tip}). However, in this paper, we focus specifically on hormesis due to its  seamless integration within the context of wireless networks. More precisely, in an antifragile system, hormesis involves a buffer utilized for both overcompensation in the system's state and preparedness for future shocks in response to randomness \cite{barbell}.
We adopt these properties by utilizing the RS groups and their interference decoding potential as a buffer. Specifically, after a blockage occurs, the network overcompensates by serving the affected user from other APs without considering interference at other users. This overcompensation is followed by dynamically modifying or creating new RS groups. However, these new groups are formed conservatively to uphold the network's flexibility for future disruptions. Additionally, after each disruption, the network adapts to the changed conditions, resulting in a different potential for interference mitigation compared to before the blockage. Consequently, beneficial RS groups can be formed during each interval, in which the network recovers from a blockage.

This work explores the fundamental concept of dynamic RS grouping as antifragile recovery method in a cell-free MIMO downlink system with perfect global instantaneous CSI. We formulate an optimization problem that jointly allocates user rates while designing common and private beamformers and RS groups to minimize the MSE of the QoS deviation. For practical implementation, we propose an outage-aware recovery framework, which decomposes the non-convex optimization problem into two solvable sub-problems.

\section{System Model}\label{ch:Sysmod}

This paper considers the RS-enabled cell-free \ac{MIMO} downlink system. More precisely, a set of single-antenna users $\mathcal{K}=\{1,\dots,K\}$ is served by a set of \acp{AP} $\mathcal{N}=\{1,\dots,N\}$, each of which equipped with $L$-antennas. The \acp{AP} are connected to a \ac{CP} with orthogonal fronthaul links of unlimited capacity, which enables central processing at the \ac{CP}. Furthermore, each user has a desired data rate $r_k^\mathsf{des}$, which represents the \ac{QoS} target of user $k$.
\subsection{Channel Model}\label{ssec:chan}
	This paper assumes quasi-static block fading channels, where the channel coefficients remain constant within the coherence time, while changing independently among coherence blocks. The channel link between \ac{AP} $n$ and user $k$ is denoted as $\vect{h}_{n,k} \in \mathbb{C}^{L\times1}$. Further, we denote the aggregate direct channel vector of user $k$ as $\vect{h}_k= [\vect{h}_{1,k}^T,\vect{h}_{2,k}^T,\dots,\vect{h}_{N,k}^T]^T \in \mathbb{C}^{NL\times 1}$ and the aggregate transmit signal vector as $\vect{x}=[\vect{x}_1^T,\vect{x}_2^T,\dots, \vect{x}_N^T]^T \in \mathbb{C}^{NL \times 1} $.

The received signal at user $k$ can be expressed as following using the aggregate vectors\vspace{-0.16cm}
\begin{align}\label{recSgn}
  y_k =& \vect{h}^H_k \vect{x} + n_k,
\end{align}
where $n_k\sim\mathcal{C}\mathcal{N}(0,\sigma_k)$ is the \ac{AWGN} sample.


\subsection{Rate Splitting}
Within the RSMA framework, user-requested messages undergo division into private and common parts. These messages are independently encoded into the private and common symbols $s_k^p$ and $s_k^c$, respectively. We assume, that the coded symbols $s_k^p$ and $s_k^c$ form an \ac{i.i.d.} Gaussian codebook. After the encoding process,the \ac{CP} shares the private (common) symbols $s_k^p$ ($s_k^c$) with a cluster of predetermined \acp{AP}, that exclusively send the beamformed private (common) symbols to user $k$. Hence, we can define the subset of users that are served by \ac{AP} $n$ with a private or common message $\mathcal{K}_n^p,\,\mathcal{K}_n^c \subseteq \mathcal{K}$, respectively, as
\begin{align}
   \mathcal{K}_n^p &= \{k\in\mathcal{K} \, | \, \text{AP } n \text{ serves } s_k^p \text{ to user }k\},\\
   \mathcal{K}_n^c &= \{k\in\mathcal{K} \, | \, \text{AP } n \text{ serves } s_k^c \text{ to user }k\}.
\end{align}
The respective beamformers $\vectw_{n,k}^p$ and $\vectw_{n,k}^c$, used by \ac{AP} $n$ to send $s_k^p$ and $s_k^c$, are created by the \ac{CP} and forwarded to \ac{AP} $n$ through the fronthaul link along with the respective private symbols $\cbrackets{s_k^p \,|\, \forall k \in \mathcal{K}_n^p}$ and common symbols $\cbrackets{s_k^c \,|\, \forall k \in \mathcal{K}_n^c}$.
After receiving the corresponding messages and beamformers,  \ac{AP} $n$  constructs the transmit signal vector $\vect{x}_n$, which is defined as
\begin{align}\label{eq:xvec}
   \vect{x}_n &= \sum_{k\in\mathcal{K}_n^p} \vectw_{n,k}^p s_k^p + \sum_{k\in\mathcal{K}_n^c} \vectw_{n,k}^c s_k^c,
\end{align}
and sends it to the users of interest. The transmit signal vector $\vect{x}_n$ is subject to the power constraint $\mathbb{E}\{\vect{x}_n^H \vect{x}_n\} \leq P^{\mathsf{Max}}_n$, or equivalently 
\begin{align} \label{powConst}
 \sum_{k\in\mathcal{K}} ||{\vect{w}_{n,k}^p}||_2^2 + \sum_{k\in\mathcal{K}} ||{\vect{w}_{n,k}^c}||_2^2 \leq P^{\mathsf{Max}}_n , \forall n \in \mathcal{N}.
\end{align}
We denote the aggregate beamforming vectors as $\vectw_k^o = \left[ (\vectw_{1,k}^o)^T,(\vectw_{2,k}^o)^T,\dots,(\vectw_{N,k}^o)^T  \right]^T  \in \mathbb{C}^{NL\times1}, \forall \, o \in \{p,c\}$ associated with $s_k^p$ and $s_k^c$, respectively. Using the above definitions, the aggregate transmit signal vector can be expressed as \begin{align}\label{eq:aggrX}
\vect{x} = \sum_{k \in \mathcal{K}_n^p}\vectw_k^p s_k^p + \sum_{k \in \mathcal{K}_n^c}\vectw_k^c s_k^c.
\end{align}

\newcommand{\RisChan}[1]{\vect{h}_{#1}}

\subsection{Achievable rates}

In this work, the \ac{SIC} scheme, adopted by the users, is utilized. We adopt a successive decoding strategy, in which user $k$ decodes the common message of other users first before decoding its own common and private message.

To this end, let $\mathcal{M}_k$ be the set of users that decode $s_k^c$, i.e.,
\begin{align}
\mathcal{M}_k = \cbrackets{j \in \mathcal{K} \,|\, \text{ user } j \text{ decodes }s_k^c}.
\end{align}
Further, let the set of users $\Phi_k$, whose common messages are decoded by user $k$, and the set of users $\Psi_k$, whose common messages are not decoded by user $k$, be defined as
\begin{align}
\Phi_k = \cbrackets{j \in \mathcal{K} \, | \, k \in\mathcal{M}_j},\, \Psi_k = \cbrackets{j \notin \mathcal{K} \, | \, k \in\mathcal{M}_j},
\end{align}
respectively. It is noted that $\Phi_k$ and $\Psi_k$ are two disjoint subsets from the set of active users $\mathcal{K}$, while we define the cardinality of $\Phi_k$ to be bounded by $D$, i.e., $|\Phi_k| \leq D.$
A decoding order is established for the sets $\Phi_k ,\, \forall k\in\mathcal{K}$, which is represented by a bijective function of the set  of $|\Phi_k|$, i.e.,
\begin{align}
   \pi_k(j) : \Phi_k \rightarrow \cbrackets{1,2,\dots,|\Phi_k|}  .
\end{align}
More precisely, assuming user $k$ decodes the common messages of user $j_1$ and user $j_2$ with $j_1 \neq j_2$, then $\pi_k(j_1) < \pi_k(j_2)$ signifies that user $k$ decodes the common message of user $j_2$ first and the common message of user $j_1$ afterwards. This implies that when user $k$ decodes the common message intended for user $j_2$, the message intended for user $j_1$ is perceived as interference. In order to capture changes in the decoding order effectively, we define the set $\Omega_{i,k}$ as
\begin{align}
   \Omega_{i,k} = \cbrackets{m\in\Phi_i \, | \, \pi_i(k) > \pi_i(m)},
\end{align}
which represents the set of users whose common messages are decoded by user $i$ after decoding the common message of user $k$.

Now, the received signal at user $k$ can be expressed as
\begin{align}\label{eq:recSig_final}
  &y_k = \overbrace{{\vecth_k^H \vectw_k^p s_k^p + \sum_{j \in \Phi_k} {\vecth_k^H \vectw_j^c s_j^c}}}^{\text{signals that are decoded}}+ \nonumber\\[-3pt]& \hspace{1.65cm}   \underbrace{\sum_{m \in \mathcal{K}\text{\textbackslash}\cbrackets{k}} \hspace{-0.35cm} {\vecth_k^H \vectw_m^p s_m^p \hspace{-0.05cm} +\hspace{-0.05cm} \sum_{\ell \in \Psi_k} {\vecth_k^H \vectw_\ell^c s_\ell^c + n_k}}}_{\text{interference plus noise}}.\\[-20pt]\nonumber
\end{align}
We denote ${\gamma_k^p}$  as the \ac{SINR} of user $k$, decoding its private message, and denote $\gamma_{i,k}^c$ as the \ac{SINR} of user $i$, decoding the common message of user $k$. Let the vector $\vectw_k = \hbrackets{ \sbrackets{\vectw_k^p}^T , \sbrackets{\vectw_k^c}^T }^T \in \mathbb{C}^{2NL\times1}$ represent the stacked private and common beamformers of user $k$ and let the vector $\vectw = \hbrackets{ (\vectw_1)^T , (\vectw_2)^T , \dots , (\vectw_K)^T}^T \in \mathbb{C}^{2KNL\times1}$ represent all beamformers.\\
With the partition in (\ref{eq:recSig_final}), the \acp{SINR} $\gamma_k^p(\vectw)$ and $\gamma_{i,k}^c(\vectw)$ can be expressed as\vspace{-0.2cm}
\begin{align}
   \gamma_k^p = \label{eq:gammap}\frac{|{{\vecth_k}^H \vectw_k^p}|^2}
                     {\sum\limits_{m \in \mathcal{K}\text{\textbackslash}\cbrackets{k}} \hspace{-0.2cm} |{{\vecth_k}^H \vectw_m^p}|^2 + \sum\limits_{\ell \in \Psi_k} |{{\vecth_k}^H \vectw_\ell^c}|^2 + \sigma^2},\\
   \gamma_{i,k}^c = \label{eq:gammac}\frac{|{{\vecth_i}^H \vectw_k^c}|^2}
                     {T_i + \sum\limits_{\ell \in \Psi_i} |{{\vecth_i}^H \vectw_\ell^c}|^2 +
                     \sum\limits_{m \in \Omega_{i,k}} |{{\vecth_i}^H \vectw_m^c}|^2},
   \\[-22pt] \nonumber
\end{align}
where $T_i = \sum_{j \in \mathcal{K}}^{\phantom{I}} |{\vecth_i}^H \vect{\omega}_j^p|^2  + \sigma^2$.
In order to assure that each user is served with its desired rate, the QoS
$r_k^{\mathsf{des}}$ of each user $k$ can be expressed as
   $r_k^p + r_k^c \geq r_k^{\mathsf{des}}$,
where $r_k^p$ is the private rate and $r_k^c$ the common rate of user $k$. The messages $s_p^k$ and $s_c^k$ are only decoded reliably if the following conditions are satisfied by the private and common rates of each user\vspace{-0.1cm}
\begin{align}\label{eq:privateRate}
B \log_2 (1+\gamma_{k}^p ) \geq& \,r_k^p , \quad \forall k \in \mathcal{K},\\
%
      B \log_2 (1+\gamma_{i,k}^c) \geq& \, r_k^c, \quad  \forall k \in \mathcal{K} ,\, \forall i \in \mathcal{M}_k,\label{eq:commonRate}
\end{align}
in which $B$ denotes the transmission bandwidth.

\subsection{Objective Performance Metric}
In this paper, a specific target throughput, determined by the \ac{QoS} requirements is considered. During the observation interval, the \ac{QoS} demands are assumed to remain constant, while the network's sum throughput is dependent on the allocated resources at time $t$. Hence, we define $r_k(t)$ to be the allocated data rate for user $k$ at time $t$. In terms of the temporal dimension, two critical points significantly influence performance behavior: $t_0$, marking the onset of the degradation, and $t_n$, indicating the point of recovery. With these markers in mind and referencing the resilience metric outlined in \cite{resiliencemetric,RobertRes,RIS_Res}, we  define the metrics for network absorption, adaptation, and time-to-recovery as follows:\vspace{-0.1cm}
	\begin{align}
	    r_\text{abs} &=  \frac{1}{K} \sum_{k\in\mathcal{K}} \frac{r_k(t_0)}{r_k^\mathsf{des}},\label{eq:rabs}\,\quad r_\text{ada} = \frac{1}{K} \sum_{k\in\mathcal{K}} \frac{r_k(t_n)}{r_k^\mathsf{des}},\quad
\end{align}\vspace{-0.3cm}
	\begin{align}
	    &\quad \quad r_\text{rec} =  \begin{cases} 1 \qquad\quad t_n-t_0\leq T_0\\  \frac{T_0}{t_n-t_0} \quad\, \text{otherwise} \end{cases}\hspace{-0.3cm}, \quad\quad\label{eq:rrec}
	\end{align}
where $T_0$ represents the desired recovery time for network operators, indicating the duration within which functional degradation remains tolerable. Combining equations (\ref{eq:rabs}) and (\ref{eq:rrec}) in a linear fashion yields the performance metric given as
	   $ r = \lambda_1 r_\text{abs} + \lambda_2 r_\text{ada} + \lambda_3 r_\text{rec}$,	where the predefined weights $\lambda_i$, $i\in\{1,2,3\}$, denote the network operator's needs regarding the robustness, the quality of adaption, or the recovery time. We define a system working in desired operation when the performance metric is one, i.e., $r=1$. To this end, we assume that $\sum_{i=1}^{3}\lambda_i = 1$.
It's worth noting that the proposed resilience metric $r$ can be temporally divided into \textit{anticipatory} actions, which encompass $r_\text{abs}$ occurring before an outage, and \textit{reactionary} actions, namely $r_\text{ada}$ and $r_\text{rec}$, unfolding after the disruption of the system. Concerning anticipatory actions, several studies focus on fortifying wireless communication networks against adverse conditions, as evidenced by works such as \cite{PrepInevit}. Thus, this paper primarily emphasizes reactionary actions, as they effectively illustrate the trade-off between solution quality and the time required for its attainment. Accordingly, we assume $\lambda_1=0$ throughout the remainder of this work.

When evaluating recovery performance, distinguishing between antifragile and resilient behaviors can prove challenging, particularly if the system can quickly return to its desired state, i.e., $r=1$, after a disruption. To highlight the disparities between antifragility and resilience, we propose that the system initially operates from a sub-optimal state, i.e., $r<1$. This approach enables differentiation between antifragility and resilience: if an increase in $r$ is evident following a disruption, antifragility is indicated. To further distinguish between these behaviors, we introduce additional volatility to the network by assuming it experiences multiple disruptions during the observation period. \vspace{-0.1cm}

\section{Problem Formulation}
To examine the recovery behavior of the dynamic \ac{RS} grouping in terms of objective performance, we address the task of minimizing the \ac{MSE} of the network-wide \ac{QoS} deviation \cite{RobertRes}. This metric aims to minimize the deviation of each user from the desired rate $r_k^\mathsf{des}$, inherently prioritizing users with larger \ac{QoS} gaps during optimization. In contrast to the (non-squared) mean absolute error discussed in \cite{RIS_Res}, the \ac{MSE} variant imposes more significant penalties for abandoning a user, meaning not serving the user at all. In such scenarios, the \ac{MSE} variant incentivizes more substantial readjustments within the system rather than abandonment. This difference is particularly crucial when a sudden disruption causes a sharp increase in the gap due to rate declines.

These rate declines can be especially pronounced if a disturbance affects a user that is assigned to decode common messages, i.e. any user $k$ with $|\mathcal{M}_k|>1$. This is due to the fact that the common rate is dictated by the lowest achievable rate among users within the group. Thus, if a disturbance affects any user decoding a common message, it can drag down the overall common rate substantially, consequently impacting the performance of all users in the group.
Hence, in this work we also aim to readjust the RS group variables after a disruption is detected within the system.

To mathematically capture the above concepts, we express the constrained squared network-wide adaption gap as:
	\begin{align}\label{Prob1}
		\underset{\vect{w},\vect{r},S}{\min} \quad & \Upxi =  \sum_{k\in \mathcal{K}} \Big|\frac{r_k^p + r_k^c}{r_k^\mathsf{des}}-1\Big|^2 \tag{P1} \\
        \text{s.t.}\quad \,\,& (\ref{powConst}), (\ref{eq:privateRate}), (\ref{eq:commonRate}), \nonumber
	\end{align}
where the stacked private and common rate of each user $k$ are denoted as $\vect{r}_k= \hbrackets{r_k^p,r_k^c}^T$ and $\vect{r}= [\vect{r}_1^T,\dots,\vect{r}_K^T]^T \in \mathbb{R}_+^K$ represents all rates.


It can be observed that the feasible set of the formulated problem is non-convex due to the non-convex nature of its constraints.
Therefore, we solve the problem efficiently by using a \ac{SCA} framework, while only redesigning the RS set variables $S$ when a disruption is detected in the system.
\subsection{Beamforming Design and Rate Allocation}
	During the design of the beamformers the RS sets $S$ are assumed to be fixed. Thus, problem (\ref{Prob1}) can be rewritten as
\begin{align}\label{Prob2}
		\underset{\vect{w},\vect{r},\vect{t}}{\min} \quad & \Upxi \quad  \quad \quad \tag{P2} \\[-5pt]
		\text{s.t.}\quad \,\, & (\ref{powConst}),\nonumber \\[-1pt]
& R_k^p - B \log_2(1+t_k^p) \leq 0, &&  \hspace{-.15cm} \forall k \in \mathcal{K}\label{eq:2.1_first}\\
    & R_k^c - B \log_2(1+t_k^c) \leq 0, &&  \hspace{-.15cm} \forall k \in \mathcal{K},\\
   &\vect{t}_k \geq 0 , &&  \hspace{-.15cm} \forall k \in \mathcal{K},\\
   &\mat{r}_k \geq 0 , &&  \hspace{-.15cm}  \forall k \in \mathcal{K},\label{eq:2.1_last}\\
   & t_k^p \leq  \gamma_k^p, &&  \hspace{-.15cm} \forall k \in \mathcal{K}, \label{eq:tkp}\\
   & t_k^c \leq  \gamma_{i,k}^c, &&  \hspace{-.15cm} \forall k \in \mathcal{K} ,\, \forall i \in \mathcal{M}_k, \label{eq:tkc}
	\end{align}
where the slack variables $\mat{R}_k= \hbrackets{R_k^p,R_k^c}^T$ and $\vect{t}_k=\hbrackets{t_k^p,t_k^c}^T$ are introduced and $\vect{t}_k  \geq  0$ and $\mat{R}_k  \geq  0$ indicates a component-wise inequality. The constraints in (\ref{eq:tkp}) and (\ref{eq:tkc}) are still non-convex but can be convexified using the \ac{SCA} approach.
Consequently, if $(\tilde{\vectw},\tilde{\vect{t}})$ is a feasible point of problem (\ref{Prob2}) then we can approximate (\ref{eq:tkp}) and (\ref{eq:tkc}) as \cite{SynBenefits}
\begin{align}
      &0 \geq \sum_{j\in\mathcal{K}\text{\textbackslash}\cbrackets{k}} |\vecth_k^H \vect{\omega}_j^p|^2 + \sum_{\ell \in \Psi_k} |\vecth_k^H \vect{\omega}_\ell^c|^2 + \sigma^2+\frac{|\vecth_k^H \overset{\phantom{I}}{\tilde{\vect{\omega}}_k^p}|^2}
      {\sbrackets{\tilde{t}_k^p}^2} \, t_k^p
      \nonumber\\
      &\hspace{1cm}-\frac{2\Re\cbrackets{ \sbrackets{\tilde{\vect{\omega}}_k^p}^H \vecth_k \vecth_k^H \vect{\omega}_k^p}}
      {\overset{}{\tilde{t}_k^p}},\forall k \in \mathcal{K}   \label{eq:approxCon1}
      \end{align}
      \begin{align}
      &0\geq {T_i}+ \sum_{\ell \in \Psi_i} |\vecth_i^H \vect{\omega}_\ell^c|^2 + \sum_{m \in \Omega_{i,k}} |\vecth_i^H \vect{\omega}_m^c|^2
      \nonumber + \frac{|\vecth_i^H \overset{\phantom{I}}{\tilde{\vect{\omega}}_k^c}|^2}
      {\sbrackets{\tilde{t}_k^c}^2} \, t_k^c\\
      & \hspace{1cm}-\frac{2\Re\cbrackets{ \sbrackets{\tilde{\vect{\omega}}_k^c}^H \vecth_i \vecth_i^H \vect{\omega}_k^c}}
      {\overset{}{\tilde{t}_k^c}},  \forall k \in \mathcal{K} ,\, \forall i \in \mathcal{M}_k ,
     \label{eq:approxCon2}
\end{align}
respectively, where $\text{Re}\cbrackets{\cdot}$ denotes the real part of a complex-valued number. Utilizing these approximations, problem (\ref{Prob2}) can be transformed into the following convex reformulation
\begin{align}\label{Prob2.1}
    \underset{\vect{w},\vect{r},\vect{t}}{\min} \quad & \Upxi \quad  \quad \quad \tag{P2.1} \\[-5pt]
		\text{s.t.}\quad \,\, & (\ref{powConst}), (\ref{eq:2.1_first}) - (\ref{eq:2.1_last}), (\ref{eq:approxCon1}) , (\ref{eq:approxCon2}), \nonumber
	\end{align}
which can be solved with conventional optimization tools. 
More precisely, we define $\mat{\Lambda}_z = [\vect{w}_z^T ,\vect{t}_z^T ]^T$ as a vector stacking the optimization variables of the beamforming design problem at iteration $z$. Similarly  $\hat{\mat{\Lambda}}_z  = [\hat{\vect{w}}_z^T ,\hat{\vect{t}}_z^T ]^T$ and $\tilde{\mat{\Lambda}}_z  = [\tilde{\vect{w}}_z^T ,\tilde{\vect{t}}_z^T ]^T$ denote the optimal solutions and the point, around which the approximations are computed, respectively. Thus, with a given point $\tilde{\mat{\Lambda}}_{z} $, an optimal solution $\hat{\mat{\Lambda}}_z $ can be obtained by solving problem (\ref{Prob2.1}). To proceed to the next iteration $z+1$, we set $\tilde{\mat{\Lambda}}_{z+1} = \hat{\mat{\Lambda}}_{z}$ and compute $\hat{\mat{\Lambda}}_{z+1}$.
\newcommand{\RisChanv}[2]{\tilde{h}_{#1,#2} + \tilde{\mat{G}}_{#1,#2}\vect{v}}

\subsection{Set design}
In this study, the \ac{RS} technique is utilized to mitigate increased interference within the network resulting from major disruptions in the system, i.e., blockages between \ac{AP} and user links. Crucial to RS effectiveness is the selection of optimal \ac{RS} sets, ensuring that the common rate $r_k^c$ for user $k$ is decodable by all users in the corresponding set $\Phi_k$. For instance, if a user $j$ within $\Phi_k$ experiences significantly lower \ac{SINR} when receiving user $k$'s common message, it impacts the transmission efficiency of the message $s_k^c$. This results in a reduction of the common rate $r_k^c$ to match user $j$'s decoding rate, thereby diminishing the effectiveness of \ac{SIC} across all users in $\Phi_k$.

The challenge is exacerbated whenever an outage occurs in the system. During such events, the existing \ac{RS} groups may become suboptimal for two main reasons: Firstly, a user's channel conditions may deteriorate to the extent that it becomes inefficient for them to remain in their current groups. Secondly, even if a user is not directly affected by any disruption, resource reallocation during the adaption process may cause changes in interference patterns. Therefore, it becomes crucial to reallocate RS groups to adapt to the altered network environment. Failure to do so can result in suboptimal performance, as existing \ac{RS} sets may no longer effectively support common message transmissions or mitigate interference.

To dynamically adjust the RS grouping process, we employ an adapted version of the dynamic RS set allocation method proposed in \cite[Algorithm 3]{SynBenefits}. The algorithm is triggered whenever a blockage occurs between AP $\tilde{n}$ and user $\tilde{k}$. It begins by removing the impacted user $\tilde{k}$ from all RS groups to mitigate potential performance degradation for other users due to a reduced common rate. Consequently, the private and common beamformers are set to zero, $\vectw_{\tilde{n},\tilde{k}}^o = \zero_L$, where $\zero_L$ denotes a column vector of length $L$ with all zero entries. To counteract the performance decline caused by the blockage, spare power at each unaffected AP is utilized, defined as $P_n^\mathsf{spare} = P^{\mathsf{Max}}_n - (\sum_{k\in\mathcal{K}} ||{\vect{w}_{n,k}^p}||_2^2 + \sum_{k\in\mathcal{K}} ||{\vect{w}_{n,k}^c}||_2^2)$.
This spare power amplifies the private beams towards the blocked user $\tilde{k}$:
\begin{align}
        \vectw_{n,\tilde{k}}^p = \vectw_{n,\tilde{k}}^p + \sqrt{P_n^\mathsf{spare}} \vectw^p_{n,\tilde{k}} / ||{\vectw^p_{n,\tilde{k}}}||^2_2, \forall n \in \mathcal{N}/\{\tilde{n}\}
\end{align}
While this amplification enhances performance for the blocked user $\tilde{k},$ it introduces more interference for other users. However, following this step, the algorithm proceeds by forming new RS groups based on the current interference scenario.

 Let $\tilde{\gamma}_{i,k}^p$ represent the SINR of user $i$ decoding the private message of user $k$ as
\begin{align}
   \tilde{\gamma}_{i,k}^p &= \frac{|{\vecth_i^H \vectw_k^p}|^2}
   {\hspace{-0.4cm}\sum\limits_{m \in \mathcal{K}\text{\textbackslash}\cbrackets{k}} \hspace{-0.35cm} |{\vecth_i^H \vectw_m^p}|^2 \hspace{-0.05cm}+\hspace{-0.2cm} \sum\limits_{\ell \in \bar{\Phi}_k} \hspace{-0.1cm} |{\vecth_i^H \vectw_\ell^c}|^2 \hspace{-0.05cm}+\hspace{-0.05cm} \sigma^2}.\nonumber\\[-12pt]\label{eq:gammap_diff}\\[-20pt]\nonumber
\end{align}
and $\tilde{\gamma}_{i,k}^{p+c} = \tilde{\gamma}_{i,k}^{p}+\gamma_{i,k}^{c}$ denote the sum-SINR of user $i$ decoding both the private and common messages of user $k$.
We define vectors $\vect{\Gamma}_k^o$ for $o \in {p,c}$, representing the ratio of each user decoding the private/common message of user $k$ to user $k$ decoding their private/common message, given as \vspace{-0.25cm}
\begin{align}\label{eq:cmdGam_p}
\vect{\Gamma}_k^p = \big[\frac{\tilde{\gamma}_{1,k}^p}{\gamma_k^p}-1 , \dots , \frac{\tilde{\gamma}_{K,k}^p}{\gamma_k^p}-1\big]^T, \\
\vect{\Gamma}_k^c = \big[\frac{{\gamma}_{1,k}^c}{\gamma_{k,k}^c}-1 , \dots , \frac{{\gamma}_{K,k}^c}{\gamma_{k,k}^c}-1\big]^T. \label{eq:cmdGam_c}
\end{align}
Additionally, $\mat{\Gamma}^o = [\vect{\Gamma}_1^o,\dots,\vect{\Gamma}_K^o]$ is a matrix comprising these vectors. Note that each ratio in $\vect{\Gamma}_k^o$ is subtracted by one. Consequently, each positive non-zero value in $\vect{\Gamma}_k^o$ represents a user, which is able to decode the respective message of user $k$ with a higher rate than user $k$ itself.


$\mat{\Gamma}^o$ is computed based on the beamforming vectors $\vectw$. Users already decoding the common message of user $k$ are disregarded by replacing the corresponding entries in $\mat{\Gamma}^o$ with an arbitrarily low value of $-\infty$. Subsequently, the highest value $\Gamma_{j,k}^o$ of $\mat{\Gamma}^o$ is identified. If this value surpasses a threshold $\epsilon^\mathsf{Pot} \in [-1,\infty]$, user $j$ is considered a potential candidate to join group $\mathcal{M}_k$ provided their decoding layers are not at full capacity, i.e., $|\Phi_j| < D$. The efficacy of this approach is confirmed by temporarily adding the user to the respective RS sets in $\mathcal{S}$, forming $\tilde{\mathcal{S}}$. To refine the quality of sets for $\{\Omega_{j,\ell}\}_{\forall \ell \in \mathcal{K}}$ within $\tilde{\mathcal{S}}$, the decoding order $\pi_j$ of user $j$ is adjusted based on the descending order of sum-SINRs, i.e., $\tilde{\gamma}_{j,\pi_j(|\Phi_j|)}^{p+c} \geq  \dots\geq\tilde{\gamma}_{j,\pi_j(2)}^{p+c}  \geq \tilde{\gamma}_{j,\pi_j(1)}^{p+c}$.

If $\overset{\phantom{.}}{\Gamma_{j,k}^o}$ belongs to $\mat{\Gamma}^p$ (i.e., $o \in {p}$), the associated beamforming vector $\vectw_k^c$ for the common message of user $k$ is updated by $\vectw_k^c = \vectw_k^c + \vectw_k^p$. If the resulting value $({{\gamma}_{j,k}^c}/{\gamma_{k,k}^c})-1$, determined with the temporary RS sets $\tilde{\mathcal{S}}$, exceeds a threshold $\epsilon^\mathsf{Val} \in [-1,\infty]$ during validation, the temporary RS sets are adopted, and corresponding values in $\mat{\Lambda}$ are updated. Otherwise, changes to $\vectw_k^c$ are disregarded, and the corresponding entry of $\mat{\Gamma}^o$ is set to $-\infty$. This process iterates until no values in $\mat{\Gamma}^o$ exceed $\epsilon^\mathsf{Pot}$.

\subsection{Outage-aware Recovery Procedure}
The recovery process can be summarized by continuously solving problem (\ref{Prob2.1}) until a blockage occurs in the system. Upon detecting the blockage, the dynamic RS grouping algorithm is activated to mitigate further performance degradation caused by unfavorable RS groups and to establish new, more beneficial groups instead. However, during the network's recovery phase, it is not just about achieving the best quality results; it is about finding the right balance between quality and time. This balance is defined by the $\lambda_i,  \, \forall i \in \{1,2,3\}$, in the resilience metric. However, every time a new RS group is formed or a user is added to an existing \ac{RS} group, the complexity of solving problem (\ref{Prob2.1}) increases. Thus, including more users in the \ac{SIC} strategy \textit{might} increase the objective performance at a \textit{certain} reduction of the recovery time. To tackle this problem, we apply a barbell strategy \cite{barbell2} to the \ac{RS} group allocation, which involves balancing a stable configuration with a riskier, potentially high-reward RS grouping. In the context of \ac{RS}, this means that every time a blockage occurs, the RS grouping algorithm is run until one user is successfully included to any \ac{RS} group. This approach maintains the performance of the remaining user rates, while enabling innovation and adaptation to changing conditions, effectively isolating performance degradation due to a bad RS group selection. Moreover, it strikes a favorable balance between efficiency and adaptability considering potential disruptions in the future. More specifically, the algorithm opts for a simpler but less efficient RS grouping at the current time to still be able to enhance the performance with RS groups if a blockage occurs in the future. In other words, if the decoding layers of every user are at full capacity after the first blockage, the algorithms flexibility in adapting to a blockage in the future is highly impeded. Therefore, we presume the initial state, prior to any blockage, to lack any RS group allocations, for the same rationale.

\section{Numerical results}
In this section, we numerically evaluate the performance of \ac{RS} as a recovery method with antifragile behaviour. To this end we assume a cell-free \ac{MIMO} system with $N=2$ \acp{AP}, each of which equipped with $L=4$ antennas. We assume the $K=12$ single-antenna users, as well as the \acp{AP}, to be distributed randomly within an area of operation, which spans $[-250,250] \times [-250,250] \text{m}^2$. For the channel links we assume Rayleigh fading channels with log-normal shadowing with 8dB standard deviation. Further, we assume a bandwidth of $B=10 \text{MHz}$, a noise power of $-100$dBm, a maximum transmit power of $P_n^\mathsf{Max} = 32$dBm per \ac{AP} and each user to require a \ac{QoS} of $r_k^\mathsf{des}=12$Mbps.
We define the occurrence of an outage as an event, where one random link between \ac{AP} $n$ and user $k$ is subject to a complete blockage. In addition, we assume the rate adaption technique in \cite[M1]{RobertRes} is utilized as a recovery mechanism right after every outage occurs, i.e., (\ref{eq:privateRate}) and (\ref{eq:commonRate}) are assumed to be feasible within the complete observation interval of 10s. Afterwards, the \ac{RS} grouping algorithm is triggered, followed by solving problem (\ref{Prob2.1}) until the next blockage occurs. The network operator's desired recovery time is set to $T_0 = 0$ms and $\lambda_2 = 1$. The dynamic RS set thresholds are set to $\epsilon^\mathsf{Pot} = -0.4$ and $\epsilon^\mathsf{Val} = -0.5$, which
means that the users, whose SINR of potentially decoding another user’s message are at least
60\% of the current SINR of decoding the message at the intended user, are considered candidates
to this user’s RS set. These users are then accepted into this set, if their SINR is still at 50\% when the SIC order at
this user is also considered. Further, the RS groups are initialized with $\Phi_k = \{k\},\forall k\in\mathcal{K}$, and the number of decoding layers at each user is set to $D=3$.

\subsection{Adaption Behaviour}
\begin{figure}
	\centering
\includegraphics[width=1.0\linewidth]{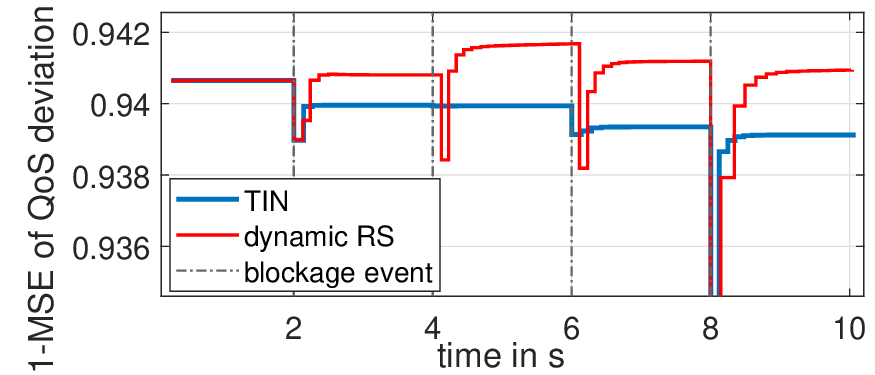}
\caption{Adaption behaviour of the TIN and dynamic RS grouping technique, in which the RS grouping technique exhibits antifragile characteristics.}
\label{adaPerf}\vspace{-0.2cm}
\end{figure}

Figure \ref{adaPerf} depicts the adaption performance of the dynamic RS grouping technique and compares it to the state-of-the-art scheme of \ac{TIN}. Note that the dynamic RS grouping is initialized without any common message decoding employed. For this reason, both methods perform identical before any blockage occurs. At the time instances [2,4,6,8]s the network experiences a blockage between one random AP-user link. After the first blockage occurs, both techniques exhibit a decline in performance. While the \ac{TIN} case demonstrates some recovery, the dynamic \ac{RS} grouping exploits the network's channel state volatility to identify advantageous RS groups. Using these groups, the performance surpasses the initial level despite encountering disruptions, which indicates antifragile behavior. The same can be observed for the next blockage at 4s. Here, the new RS groups cause the beamformers to significantly reorganize. This results in a short dip in performance, due to the \acp{SCA} converging to a new local optimum. However, the performance in this new optimum exceeds the performance of the previous state again. In contrast, the \ac{TIN} method does not show the dip in performance because it converges towards the same local minimum. In turn, the \ac{TIN} method does not recover by exceeding its previous performance. Up until this point, the dynamic RS grouping has exhibited antifragile characteristics. After the third blockage at 6s, however,  the cumulative stress on the network, caused by the blockage events, reaches a "tipping point", after which the system begins to transition into a more fragile state \cite{Antifrag_Tip}. The figure illustrates that the system's adaptation following the blockages at 6s and 8s still yields favorable performance when compared to the TIN case. However, the performance no longer surpasses the performance level observed before the blockage events. Interestingly, the RS grouping technique provides better performance at the end of the observed interval compared to its initial state. Additionally, the figure indicates a slowdown of the dynamic RS grouping in convergence performance after each blockage, attributed to the increased number of RS groups in the system.

\begin{figure}
	\centering
	\includegraphics[width=1\linewidth]{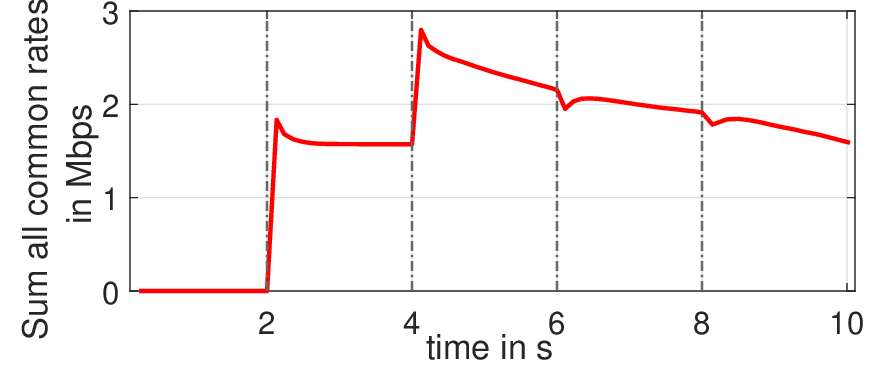}
	\caption{Sum of all common rates during the observed time interval.}
	\label{lambVals}\vspace{-0.2cm}
\end{figure}
\subsection{Impact of Rate Splitting}
The sum of all common rates within the observed time interval are depicted in Figure \ref{lambVals}. The figure shows the effect of overcompensation after the RS grouping algorithm reacts to the outage, depicted by the small peaks after the first two blockages. Interestingly, it is after these blockages, where the RS grouping also exhibits an antifragile behaviour. The last two blockages are characterized by a negative peak, in which the blockage-affected users might have been removed from already existing RS groups, resulting in continuously declining common rates. It is also for these two blockages that the network started to exhibit a more fragile behaviour.

\section{Conclusion}\label{ch:conc}
This work explores the concept of dynamic \ac{RS} grouping as a means to enhance the resilience of cell-free MIMO downlink systems in the face of disruptions. By leveraging the flexibility of RS grouping, the system can adapt to changing network conditions and mitigate the impact of blockages, ultimately improving overall performance. Through numerical simulations, we have demonstrated the effectiveness of the proposed approach in recovering from disruptions and even surpassing initial performance levels, showcasing its antifragile behavior.

\balance
%




%







\footnotesize
\bibliographystyle{IEEEtran}
\bibliography{references}
\balance
\begin{acronym}
\setlength{\itemsep}{0.1em}
\acro{AP}{access point}
\acro{AF}{amplify-and-forward}
\acro{AWGN}{additive white Gaussian noise}
\acro{B5G}{Beyond 5G}
\acro{BS}{base station}
\acro{CB}{coherence block}
\acro{CE}{channel estimation}
\acro{C-RAN}{Cloud Radio Access Network}
\acro{CMD}{common message decoding}
\acro{CP}{central processor}
\acro{CSI}{channel state information}
\acro{CRLB}{Cramér-Rao lower bound}
\acro{D2D}{device-to-device}
\acro{DC}{difference-of-convex}
\acro{DFT}{discrete Fourier transformation}
\acro{DL}{downlink}
\acro{GDoF}{generalized degrees of freedom}
\acro{IC}{interference channel}
\acro{i.i.d.}{independent and identically distributed}
\acro{IRS}{intelligent reflecting surface}
\acro{IoT}{Internet of Things}
\acro{LoS}{line-of-sight}
\acro{LSF}{large scale fading}
\acro{M2M}{Machine to Machine}
\acro{MISO}{multiple-input and single-output}
\acro{MIMO}{multiple-input and multiple-output}
\acro{MRT}{maximum ratio transmission}
\acro{MRC}{maximum ratio combining}
\acro{MSE}{mean square error}
\acro{NOMA}{non-orthogonal multiple access}
\acro{NLoS}{non-line-of-sight}
\acro{PSD}{positive semidefinite}
\acro{QCQP}{quadratically constrained quadratic programming}
\acro{QoS}{quality-of-service}
\acro{RF}{radio frequency}
\acro{RC}{reflect coefficient}
\acro{RIS}{reconfigurable intelligent surface}
\acro{RS-CMD}{rate splitting and common message decoding}
\acro{RSMA}{rate-splitting multiple access}
\acro{RS}{rate splitting}
\acro{SCA}{successive convex approximation}
\acro{SDP}{semidefinite programming}
\acro{SDR}{semidefinite relaxation}
\acro{SIC}{successive interference cancellation}
\acro{SINR}{signal-to-interference-plus-noise ratio}
\acro{SOCP}{second-order cone program}
\acro{SVD}{singular value decomposition }
\acro{TIN}{treating interference as noise}
\acro{TDD}{time-division duplexing}
\acro{TSM}{topological signal management}
\acro{UHDV}{Ultra High Definition Video}
\acro{UL}{uplink}

\acro{AF}{amplify-and-forward}
\acro{AWGN}{additive white Gaussian noise}
\acro{B5G}{Beyond 5G}
\acro{BS}{base station}
\acro{C-RAN}{Cloud Radio Access Network}
\acro{CSI}{channel state information}
\acro{CMD}{common-message-decoding}
\acro{CM}{common-message}
\acro{CoMP}{coordinated multi-point}
\acro{CP}{central processor}
\acro{D2D}{device-to-device}
\acro{DC}{difference-of-convex}
\acro{EE}{energy efficiency}
\acro{IC}{interference channel}
\acro{i.i.d.}{independent and identically distributed}
\acro{IRS}{intelligent reflecting surface}
\acro{IoT}{Internet of Things}
\acro{LoS}{line-of-sight}
\acro{LoSC}{level of supportive connectivity}
\acro{M2M}{Machine to Machine}
\acro{NOMA}{non-orthogonal multiple access}
\acro{MISO}{multiple-input and single-output}
\acro{MIMO}{multiple-input and multiple-output}
\acro{MMSE}{minimum mean squared error}
\acro{MRT}{maximum ratio transmission}
\acro{MRC}{maximum ratio combining}
\acro{NLoS}{non-line-of-sight}
\acro{PA}{power amplifier}
\acro{PSD}{positive semidefinite}
\acro{QCQP}{quadratically constrained quadratic programming}
\acro{QoS}{quality-of-service}
\acro{RF}{radio frequency}
\acro{RRU}{remote radio unit}
\acro{RS-CMD}{rate splitting and common message decoding}
\acro{RS}{rate splitting}
\acro{SDP}{semidefinite programming}
\acro{SDR}{semidefinite relaxation}
\acro{SIC}{successive interference cancellation}
\acro{SCA}{successive convex approximation}
\acro{SINR}{signal-to-interference-plus-noise ratio}
\acro{SOCP}{second-order cone program}
\acro{SVD}{singular value decomposition }
\acro{TP}{transition point}
\acro{TIN}{treating interference as noise}
\acro{UHDV}{Ultra High Definition Video}
\acro{URLLC}{ultra reliable and low-latency communication}
\acro{LoSC}{level of supportive connectivity}
\end{acronym}

\balance
\end{document}